\documentclass[]{krakproc}
\usepackage{graphicx}
\begin{document}

\title{MHD Turbulence: Properties of Alfven, Slow and Fast Modes}            
\author{A. Lazarian and A. Beresnyak}
\institute{University of Wisconsin-Madison, Dept. of Astronomy}
\markboth{A. Lazarian and A. Beresnyak}{MHD Turbulence}

\maketitle

\begin{abstract} 
We summarise basic properties of MHD turbulence.
 First,  MHD turbulence is not so messy
as it is believed. In fact, the 
notion of strong non-linear coupling of compressible and 
incompressible motions along MHD cascade is not tenable. Alfven, slow 
and fast modes of
MHD turbulence follow their own cascades and exhibit degrees of anisotropy
consistent with theoretical expectations. Second, the fast decay of 
turbulence
is not related to the compressibility of fluid. Rates of decay of 
compressible and incompressible motions are very similar.
Third, the properties of Alfven and slow modes are similar to
their counterparts in the incompressible MHD. The properties
of fast modes are similar to accoustic turbulence, which does
 require more studies. Fourth, the density at low Mach numbers
and logarithm of density at higher Mach numbers exhibit
Kolmogorov-type spectrum.

\end{abstract}

\section{What is MHD Turbulence?}

A fluid of viscosity $\nu$ becomes turbulent when the rate of viscous 
dissipation, which is  $\sim \nu/L^2$ at the energy injection scale $L$, 
is much smaller than
the energy transfer rate $\sim V_L/L$, where $V_L$ is the velocity dispersion
at the scale $L$. The ratio of the two rates is the Reynolds number 
$Re=V_LL/\nu$. In general, when $Re$ is larger than $10-100$
the system becomes turbulent. Chaotic structures develop gradually as 
$Re$ increases,
and those with $Re\sim10^3$ are appreciably less chaotic than those
with $Re\sim10^8$. Observed features such as star forming clouds are
very chaotic for $Re>10^8$. 
This makes it difficult to simulate realistic turbulence. 
The currently available
3D simulations containing 512 grid cells along each side
can have $Re$ up to $\sim O(10^3)$
and are limited by their grid sizes. 
Therefore, it is essential to find ``{\it scaling laws}" in order to
extrapolate numerical calculations ($Re \sim O(10^3)$) to
real astrophysical fluids ($Re>10^8$). 
We show below that even with its limited resolution, numerics is a great 
tool for {\it testing} scaling laws.

Kolmogorov theory provides a scaling law for {\it incompressible} 
{\it non}-magnetized hydrodynamic turbulence (Kolmogorov 1941).
This law provides a statistical relation
between the relative velocity $v_l$ of fluid elements and their separation
$l$, namely, $v_l\sim l^{1/3}$.  An equivalent description is to 
express spectrum $E(k)$
as a function of wave number $k$ ($\sim 1/l$).
The two descriptions are related by $kE(k) \sim v_l^2$. The famous
Kolmogorov spectrum is  $E(k)\sim k^{-5/3}$. The applications of 
Kolmogorov theory range from engineering research to
meteorology (see Monin \& Yaglom 1975) but its astrophysical
applications are poorly justified and the application
of the Kolmogorov theory can lead to erroneous conclusions
(see reviews by Lazarian et al.
2003 and Lazarian \& Yan 2003).

Let us consider {\it incompressible} MHD turbulence 
first\footnote{Traditionally there is insufficient interaction between
researchers dealing with {\it compressible} and {\it incompressible}
MHD turbulence. 
This is very unfortunate, as we will show later that there are many 
similarities between the properties 
of incompressible MHD turbulence and those of its compressible counterpart.}.
There have long been understanding that the MHD turbulence
is anisotropic
(e.g. Shebalin et al.~1983). Substantial progress has been achieved
recently by Goldreich \& Sridhar (1995; hereafter GS95), who made an
ingenious prediction regarding relative motions parallel and
perpendicular to magnetic field {\bf B} for incompressible
MHD turbulence. 
An important observation that leads to understanding of the GS95
scaling\footnote{Here we provide a more intuitive description, while
a GS95 presents a more mathematical one.} is that magnetic field 
cannot prevent mixing motions
of magnetic field lines if the motions
are perpendicular to the magnetic field. Those motions will cause, however,
waves that will propagate along magnetic field lines.
If that is the case, 
the time scale of the wave-like motions along the field, 
i.e. $\sim l_{\|}/V_A$,
($l_{\|}$ is the characteristic size of the perturbation along 
the magnetic field and 
$V_A=B/\sqrt{4 \pi \rho}$ is 
the local Alfven speed) will be equal to the hydrodynamic time-scale, 
$l_{\perp}/v_l$, 
where $l_{\perp}$ is the characteristic size of the perturbation
perpendicular to the magnetic field.
The mixing motions are 
hydrodynamic-like\footnote{Simulations in Cho, Lazarian \& Vishniac
((2002a, 2003b) that the mixing motions are hydrodynamic up to high order. 
These motions according to Cho et al. (2003) allow efficient turbulent
heat conduction.}.
They obey Kolmogorov scaling,
$v_l\propto l_{\perp}^{1/3}$,  because incompressible turbulence is assumed.
Combining the two relations above
we can get the GS95 anisotropy, $l_{\|}\propto l_{\perp}^{2/3}$
(or $k_{\|}\propto k_{\perp}^{2/3}$ in terms of wave-numbers).
If  we interpret $l_{\|}$ as the eddy size in the direction of the 
local 
magnetic field.
and $l_{\perp}$ as that in the perpendicular directions,
the relation implies that smaller eddies are more elongated.
The latter is natural as it the energy in hydrodynamic motions
decreases with the decrease of the scale. As the result it gets more
and more difficult for feeble hydrodynamic motions to bend magnetic 
field lines. 

GS95 predictions have been confirmed 
numerically (Cho \& Vishniac 2000; Maron \& Goldreich 2001;
Cho, Lazarian \& Vishniac 2002a, hereafter CLV02a; see also CLV03a); 
they are in good agreement with observed and inferred astrophysical spectra 
(see CLV03a). What happens in a compressible MHD? Does any part
of GS95 model survives?
Literature on the properties of compressible MHD is very rich (see reviews
by Pouquet 1999; Cho \& Lazarian 2003b and references therein).
Higdon (1984) theoretically studied density fluctuations
in the interstellar MHD turbulence.
Matthaeus \& Brown (1988) studied nearly incompressible MHD at low Mach
number and Zank \& Matthaeus (1993) extended it. In an important paper
Matthaeus et al.~(1996) numerically
explored anisotropy of compressible MHD turbulence. However, those
papers do not provide universal scalings of the GS95 type.

The complexity of the
compressible magnetized turbulence with magnetic field made some
researchers believe that the phenomenon is too complex to expect any
universal scalings for molecular cloud research.
Alleged high coupling of compressible
and incompressible motions is often quoted to justify this 
point of view (see discussion of this point below).

In what follows we discuss the turbulence in the presence of
regular magnetic field which is  comparable
to the fluctuating one. Therefore 
for most part of our discussion, we shall discuss results
obtained for 
$\delta V \sim \delta B/\sqrt{4 \pi \rho} \sim B_0/\sqrt{4 \pi \rho}$,
where $\delta B$ is the r.m.s. strength of the random magnetic field.
However, we would argue that our choice is not so restrictive as
it may be seen. Indeed, at the scales where the
 velocity perturbations are much larger
than the Alfven velocity, the dynamical importance of magnetic field
is small. Therefore we expect that at those scales turbulent motions 
are close to hydrodynamic ones. At smaller scales where the local
turbulent velocity gets smaller than the Alfven speed we believe that
our picture will be approximately true. We think that the local
magnetic field should act as $B_0$, while the small scale perturbations
 happen in respect to that local field. This reasoning is in agreement with
calculations in Cho, Lazarian \& Vishniac (2003b) and Cho \& Lazarian (2003a).

\section{Does the Decay of MHD Turbulence Depend on Compressibility? }

Many astrophysical problems, e.g. the 
turbulent support of molecular clouds (see review by McKee 1999), critically
depends on the rate of turbulence decay. 
For a long time magnetic fields were thought to be the means of
making turbulence less dissipative.
Therefore it came as a surprise when numerical calculations by
Mac Low et al. (1998) and Stone, Ostriker, \& Gammie (1998) indicated that 
compressible MHD turbulence
decays as fast as the hydrodynamic turbulence. This gives rise to a 
erroneous belief that it is the compressibility that 
is responsible for the rapid decay of MHD turbulence.  

This point of view has been  challenged in 
Cho \& Lazarian (2002, 2003a, henceforth CL02 and
CL03, respectively).  
In these papers a
 technique of separating different MHD modes was developed and used
(see Fig.~1). 
This allowed us to follow how the energy was redistributed between
these modes.
\begin{figure*}
  \includegraphics[width=0.90\textwidth]{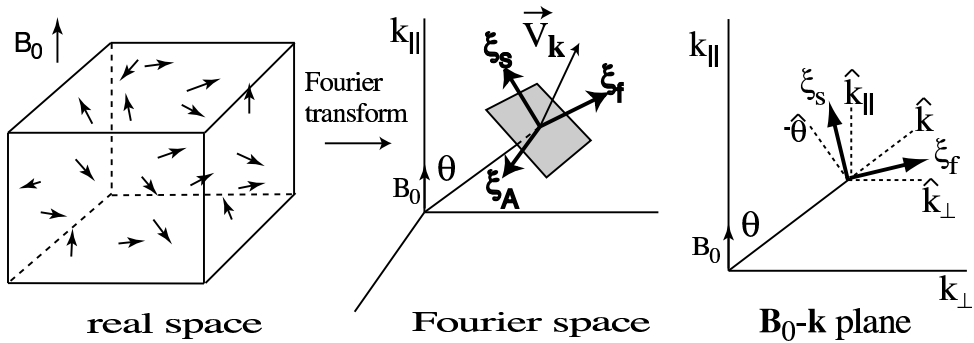}
  \caption{
      Separation method. We separate Alfven, slow, and fast modes in Fourier
      space by projecting the velocity Fourier component ${\bf v_k}$ onto
      bases ${\bf \xi}_A$, ${\bf \xi}_s$, and ${\bf \xi}_f$, respectively.
      Note that ${\bf \xi}_A = -\hat{\bf \varphi}$. 
      Slow basis ${\bf \xi}_s$ and fast basis ${\bf \xi}_f$ lie in the
      plane defined by ${\bf B}_0$ and ${\bf k}$.
      Slow basis ${\bf \xi}_s$ lies between $-\hat{\bf \theta}$ and 
      $\hat{\bf k}_{\|}$.
      Fast basis ${\bf \xi}_f$ lies between $\hat{\bf k}$ and 
      $\hat{\bf k}_{\perp}$. 
}
\label{fig_separation}
\end{figure*}

Should the different MHD modes be strongly coupled
when the turbulence is strong? A {\it naive}  answer is ``yes''.
Indeed, strong turbulence implies strong field line wondering.
This mixes up Alfven and fast modes. In addition, one can show
through calculations that the magnetic non-linearities result
in the drainage of energy from Alfvenic cascade. However,
a remarkable feature of the GS95 model is that
Alfven perturbations cascade to small scales over just one wave
period, which gets shorter and shorter as we move along the
cascade. The competing effects coupling different modes usually
require more time\footnote{This reasoning shows that at the
energy injection scale when $\delta B\sim B_0$ the coupling 
between the modes is appreciable.}.
We note that as the consequence of this reasoning we should 
assume that the properties of the Alfvenic
cascade  (incompressible cascade!) should not  strongly depend on 
the sonic Mach number.

Are the arguments above correct?
The generation of compressible motions 
(i.e. {\it radial} components in Fourier space) from Alfvenic turbulence
is a measure of mode coupling.
How much energy in compressible motions is drained from Alfvenic cascade?
According to closure calculations (Bertoglio, 
Bataille, \& Marion 2001; see also Zank \& Matthaeus 1993),
the energy in compressible modes in {\it hydrodynamic} turbulence scales
as $\sim M_s^2$ if $M_s<1$.
CL03 conjectured that this relation can be extended to MHD turbulence
if, instead of $M_s^2$, we use
$\sim (\delta V)_{A}^2/(a^2+V_A^2)$. 
(Hereinafter, we define $V_A\equiv B_0/\sqrt{4\pi\rho}$, where
$B_0$ is the mean magnetic field strength.) 
However, since the Alfven modes 
are anisotropic, 
this formula may require an additional factor.
The compressible modes are generated inside the so-called
Goldreich-Sridhar cone, which takes up $\sim (\delta V)_A/ V_A$ of
the wave vector space. The ratio of compressible to Alfvenic energy 
inside this cone is the ratio given above. 
If the generated fast modes become
isotropic (see below), the diffusion or, ``isotropization'' of the
fast wave energy in the wave vector space increase their energy by
a factor of $\sim V_A/(\delta V)_A$. This  results in
\begin{equation}
  \frac{ (\delta V)_{rad}^2 }{ (\delta V)_A^2 }   \sim
 \left[ \frac{ V_A^2 + a^2 }{ (\delta V)^2_A } 
        \frac{ (\delta V)_A }{ V_A }   \right]^{-1},
\label{eq_high2}
\end{equation}
where $(\delta V)_{rad}^2$ and $(\delta V)_{A}^2$ are energy
of compressible  and Alfven modes, respectively.
Eq.~(\ref{eq_high2}) suggests that the drain of energy from
Alfvenic modes is marginal along the cascade\footnote{
	The marginal generation of compressible 
        modes is in agreement with 
        earlier studies by Boldyrev, Nordlund, \& Padoan (2002) and 
        Porter, Pouquet, \& Woodward (2002),
        where the
        velocity was decomposed into a potential component
        and a solenoidal component. A recent study by
	Vestuto, Ostriker \& Stone 
(2003) is also consistent with this conclusion. }
when the amplitudes of perturbations
are weak. Results of calculations
shown in Fig.~2 support the theoretical predictions.

We may summarize this issue in the following way. For the incompressible
motions to decay fast, there is no requirement of coupling with 
compressible motions\footnote{
   The reported (see Mac Low et al.~1998) decay of the {\it total}
   energy of turbulent motions $E_{tot}$ follows $t^{-1}$ which can
   be understood if we account for the fact that the energy is being
   injected at the scale smaller than the scale of the system. Therefore
   some energy originally diffuses to larger scales through the inverse 
   cascade. Our calculations (Cho \& Lazarian, unpublished), 
   stimulated by illuminating 
   discussions with Chris McKee, show that if this energy 
   transfer is artificially
   prevented by injecting the energy on the scale of the computational box, 
   the scaling of $E_{tot}$  becomes closer to $t^{-2}$.}. 
The marginal coupling of the compressible 
and incompressible modes allows us to study these modes
separately. 
 
\begin{figure*}
  \includegraphics[width=0.34\textwidth]{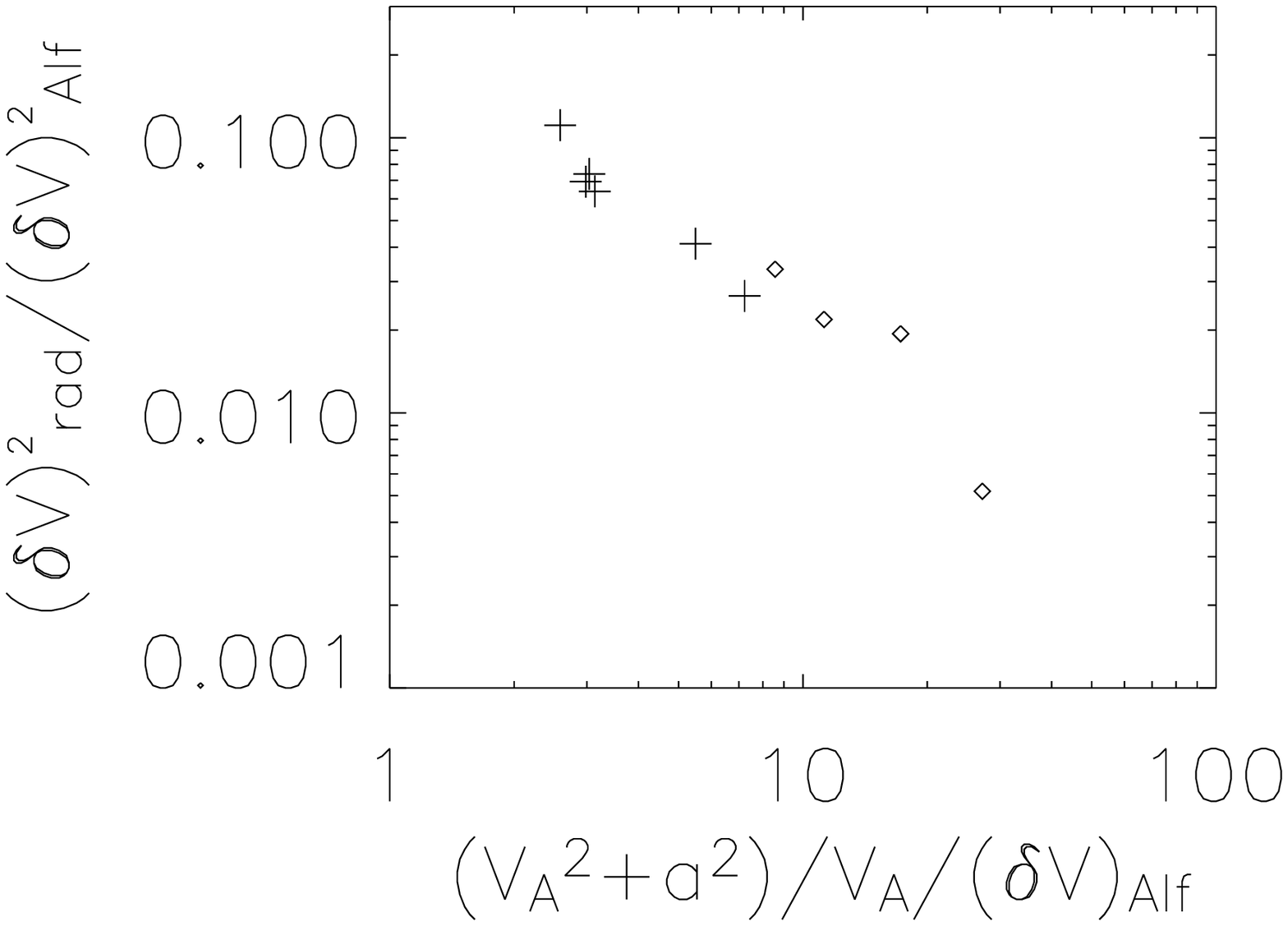}
\hfill
  \includegraphics[width=0.24\textwidth]{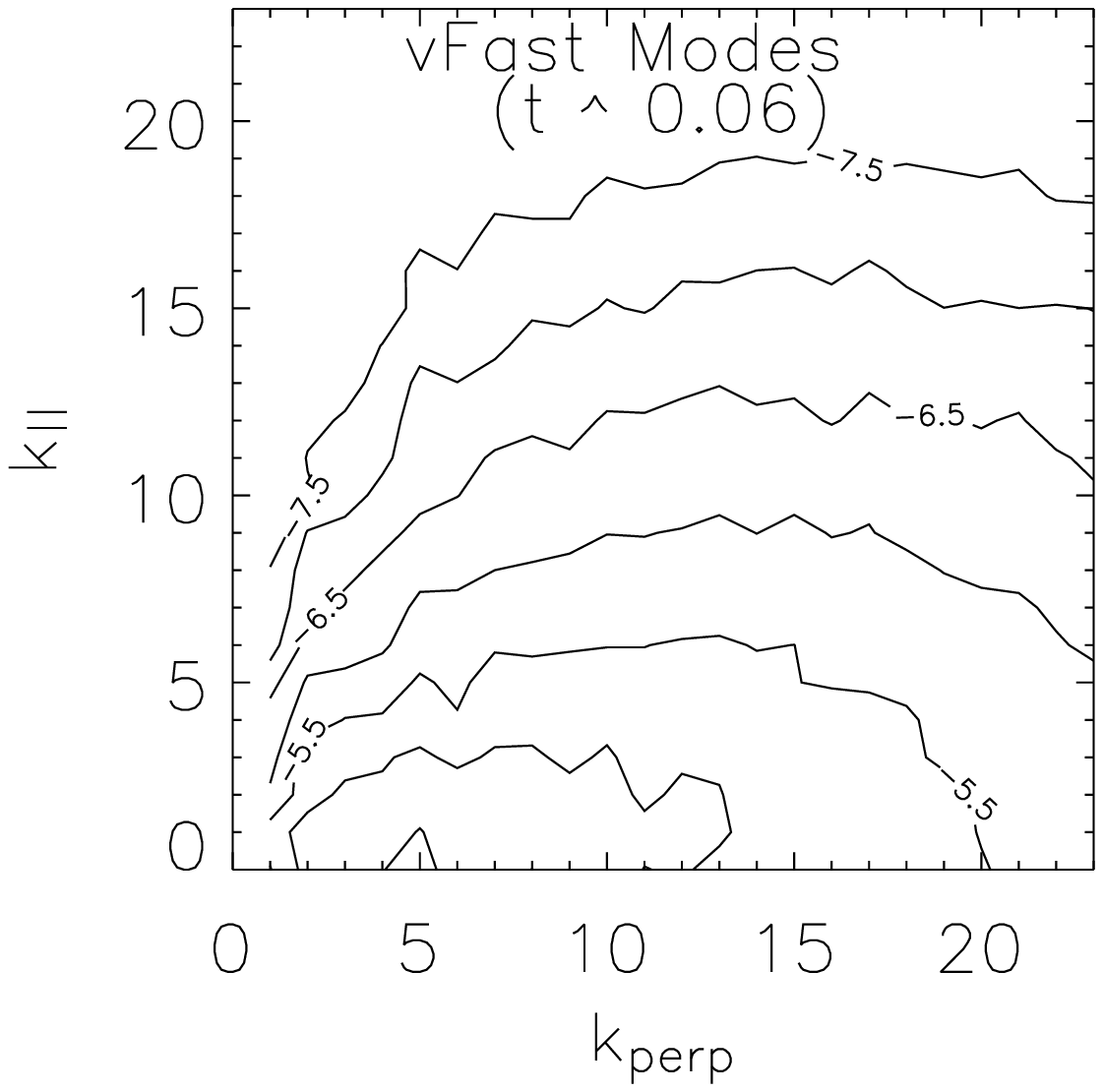}
\hfill
  \includegraphics[width=0.3\textwidth]{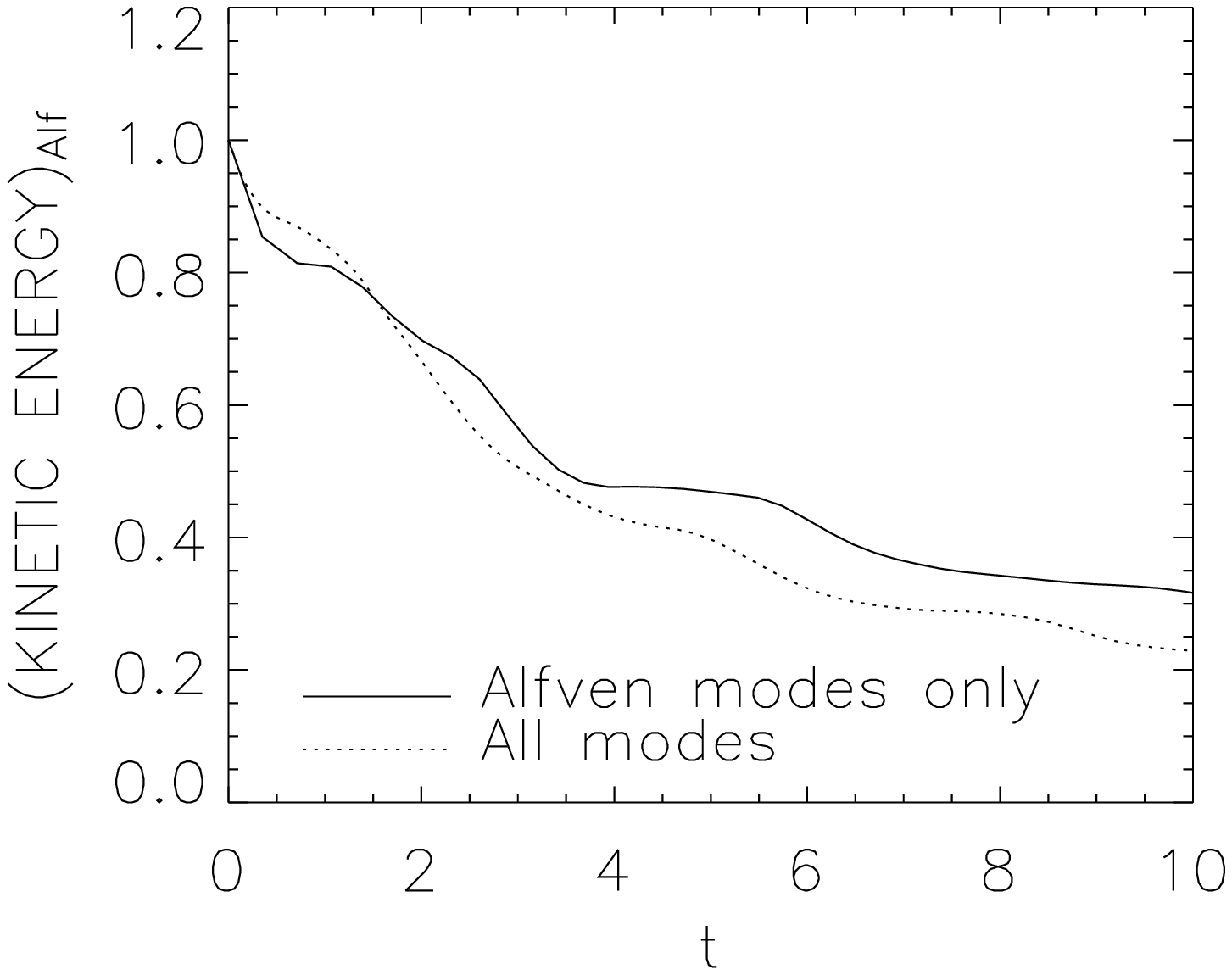}
  \caption{
      Mode coupling studies.
    (a){\it left:}  Square of the r.m.s. velocity of the compressible modes.
        We use $144^3$ grid points. Only Alfven modes are allowed
        as the initial condition.
        ``Pluses'' are for low $\beta$ cases ($0.02 \leq \beta \leq 0.4$).
        ``Diamonds'' are for high  $\beta$ cases ($1 \leq \beta \leq 20$).
    (b){\it middle:} Generation of fast modes. Snapshot is taken at t=0.06 from
        a simulation (with $144^3$ grid points) 
        that started off with Alfven modes only.
        Initially, $\beta$ (ratio of gas to magnetic pressure, $P_g/P_{mag}$) 
          $=0.2$ and 
          $M_s$ (sonic Mach number) $\sim 1.6$.
    (c){\it right:} Comparison of decay rates.
        Decay of Alfven modes is not much affected by other 
       (slow and fast) modes. We use $216^3$ grid points.
        Initially, $\beta=0.02$ and 
        $M_s\sim 4.5$ for the solid line and 
        $M_s\sim 7$ for the dotted line. 
        Note that initial data are, in some sense, identical for
        the solid and the dotted lines.
        The sonic Mach number for the solid line is smaller
        because we removed fast and slow modes from the initial data before
        the decay simulation.
        For the dotted line, we did {\it not} remove any modes from the
        initial data. From CL03.
}
\label{fig_coupling}
\end{figure*}

\section{What are the scalings for velocity and magnetic field?}

Some hints about effects of compressibility can be inferred from 
the GS95 seminal paper. More discussion was
presented in Lithwick \& Goldreich (2001), which primary deals with electron
density fluctuations in the regime of high  $\beta$, i.e.
($\beta\equiv P_{gas}/P_{mag}\gg 1$). 
As the incompressible regime 
corresponds to $\beta\rightarrow \infty$, so it is natural
to expect that for $\beta\gg 1$ the GS95 picture would
persist. Lithwick \&
Goldreich (2001) also speculated that for low $\beta$ plasmas the GS95
scaling of slow modes may be applicable. 
A detailed 
study of compressible mode scalings  is given in CL02 and CL03. 

Our considerations above about the mode coupling can guide us
in the discussion below. Indeed,
if Alfven cascade evolves on its own, it is natural to assume that 
slow modes exhibit the GS95 scaling.
Indeed, slow modes in gas 
pressure dominated environment (high $\beta$ plasmas) are
similar to the pseudo-Alfven modes in incompressible regime 
(see GS95; Lithwick \& Goldreich 2001). The latter modes do follow
the GS95 scaling. 
In magnetic pressure dominated environments  or low $\beta$ plasmas, 
slow modes are density perturbations propagating with the
sound speed $a$ parallel to the mean magnetic field. 
Those perturbations are essentially
static for $a\ll V_A$. 
Therefore Alfvenic turbulence is expected to mix density
perturbations as if they were passive scalar. This also induces the
GS95 spectrum.

The fast waves in low $\beta$ regime propagate at $V_A$ irrespectively
of the magnetic field direction. 
In high $\beta$ regime, the properties of fast modes are similar, 
but the propagation speed is the sound speed $a$.
Thus the mixing motions induced by Alfven waves should marginally
affect the fast wave
cascade. It is expected to
be analogous to the acoustic wave cascade and hence be isotropic.

Results of numerical calculations from Cho \& Lazarian (CL03, Cho \&
Lazarian 2004) 
for magnetically dominated media similar to that in molecular 
clouds are
shown in Fig.~3. They support theoretical considerations above. 

\begin{figure*}
  \includegraphics[width=0.95\textwidth]{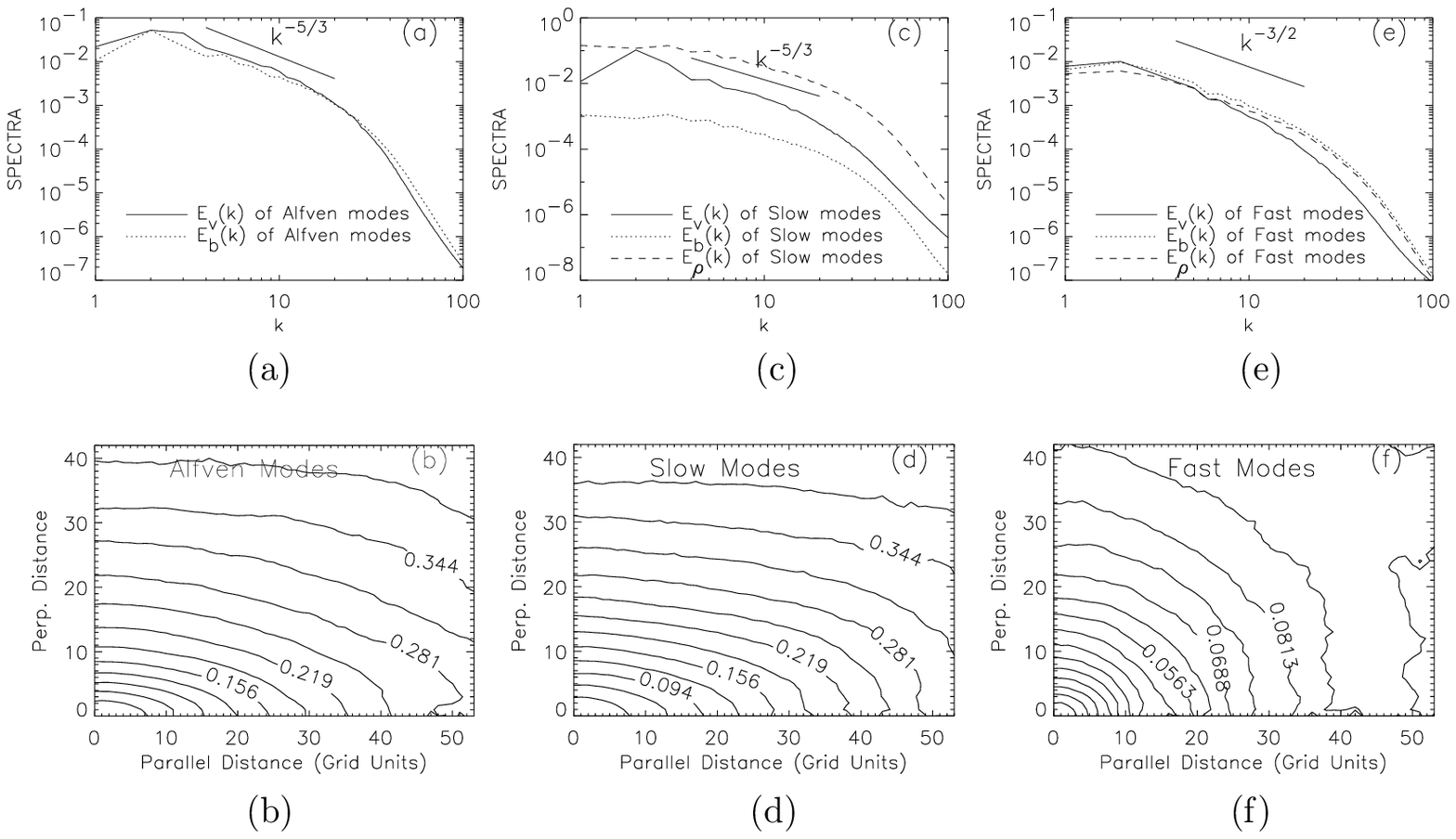}

  \caption{          $M_s\sim 2.2$, $M_A\sim 0.7$, $\beta\sim 0.2$,
           and $216^3$ grid points.
          (a) Spectra of Alfv\'en modes follow a Kolmogorov-like power
              law.
          (b) Eddy shapes
              (contours of same second-order structure function, $SF_2$)
              for velocity of Alfv\'en modes
              shows anisotropy similar to the GS95
         ($r_{\|}\propto r_{\perp}^{2/3}$ or $k_{\|}\propto
                                              k_{\perp}^{2/3}$).
              The structure functions are measured in directions
              perpendicular or
              parallel to the local mean magnetic field in real space.
              We obtain real-space velocity and magnetic fields
              by inverse Fourier transform of
              the projected fields.
          (c) Spectra of slow modes also follow a Kolmogorov-like power
              law.
          (d) Slow mode velocity shows anisotropy similar to the GS95.
              We obtain contours of equal $SF_2$ directly in real space
              without going through the projection method,
              assuming slow mode velocity is nearly parallel to local
              mean magnetic field in low $\beta$ plasmas.
          (e) Spectra of fast modes are compatible with
              the IK spectrum.
          (f) The magnetic $SF_2$ of
              fast modes shows isotropy.  From CL02
    } 
\label{fig_M2}
  \vfill
  \vskip 1.0cm
\hskip 0.02\textwidth
    \includegraphics[width=0.29\textwidth]{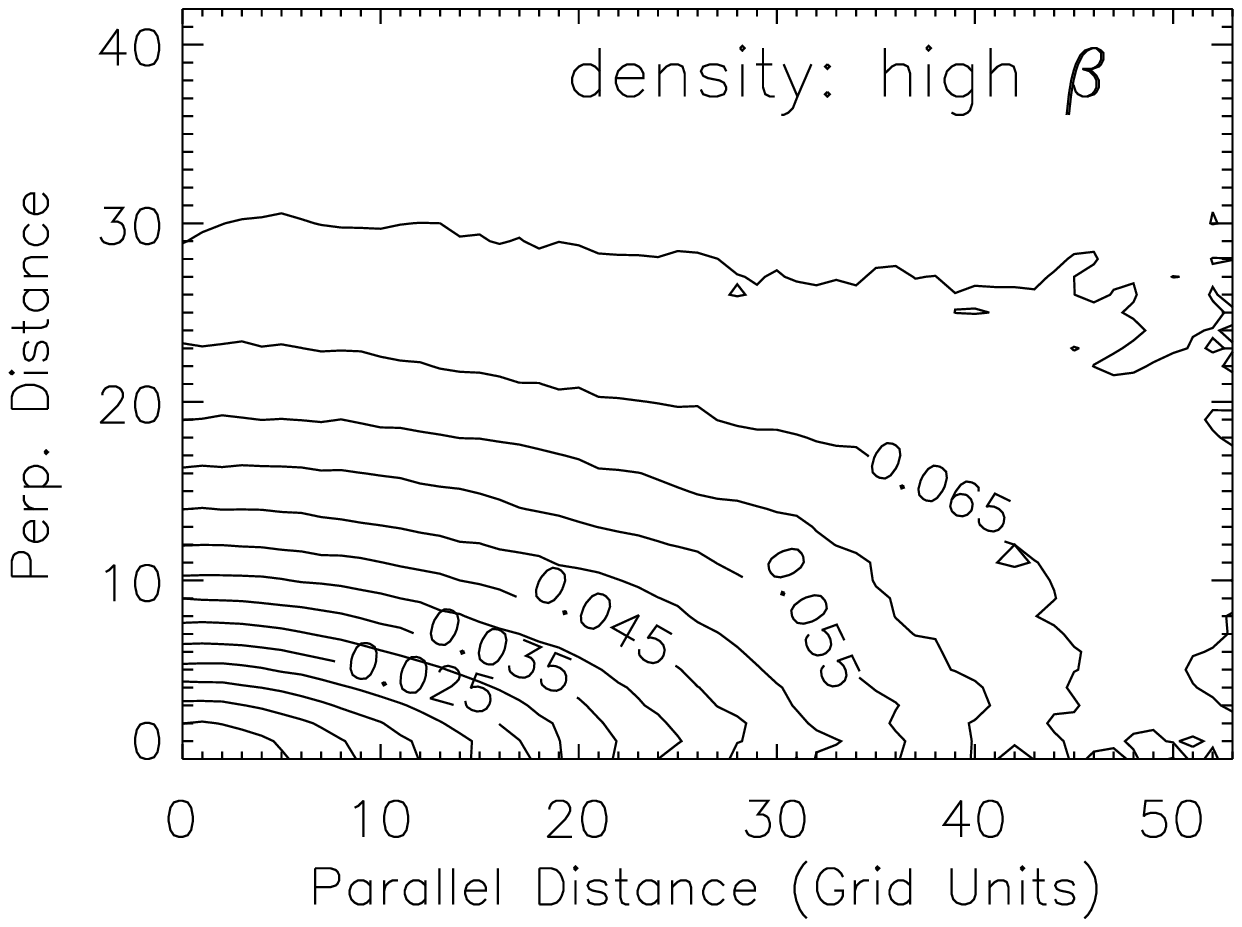}
\hfill
  \includegraphics[width=0.29\textwidth]{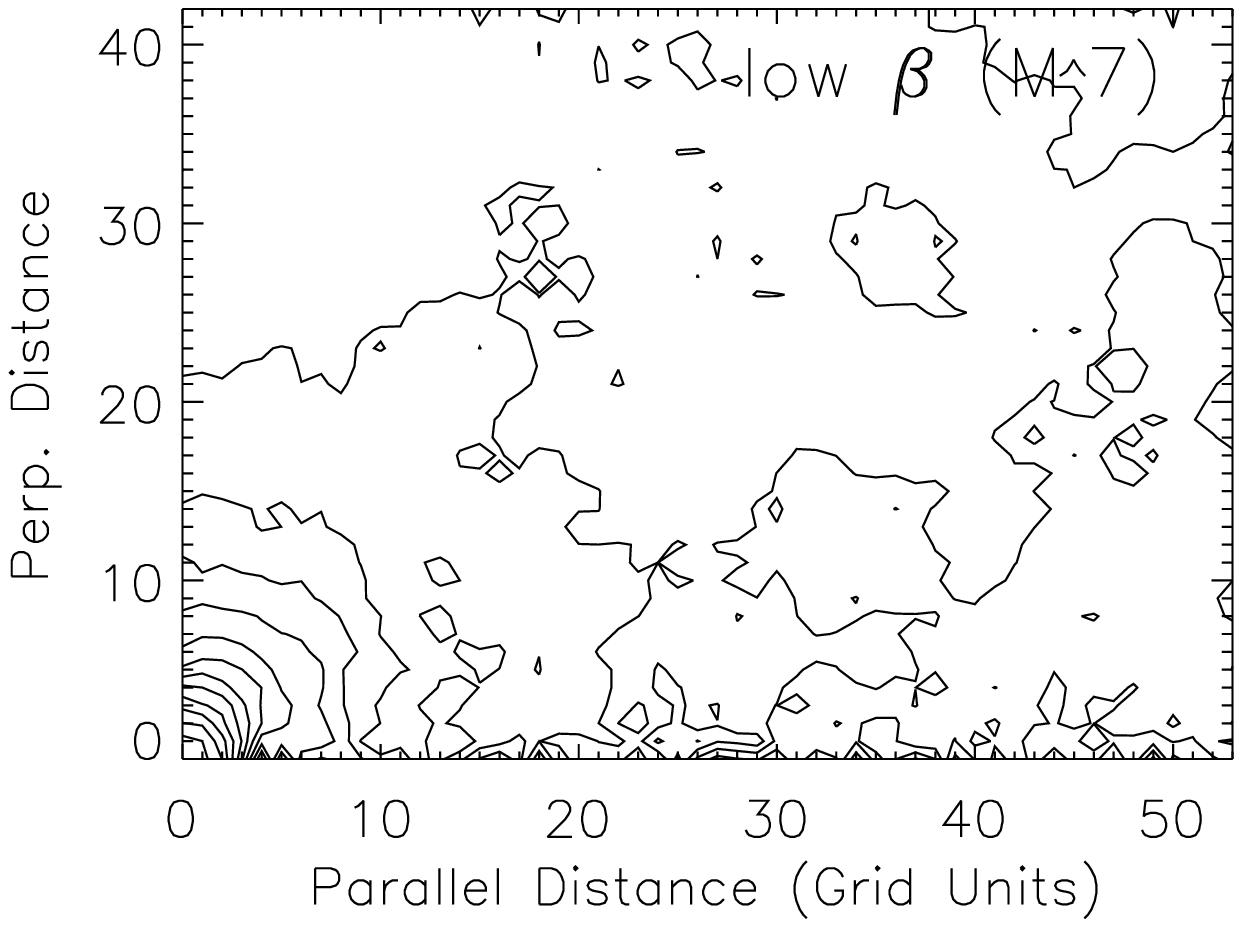}
\hfill
\includegraphics[width=0.32\textwidth]{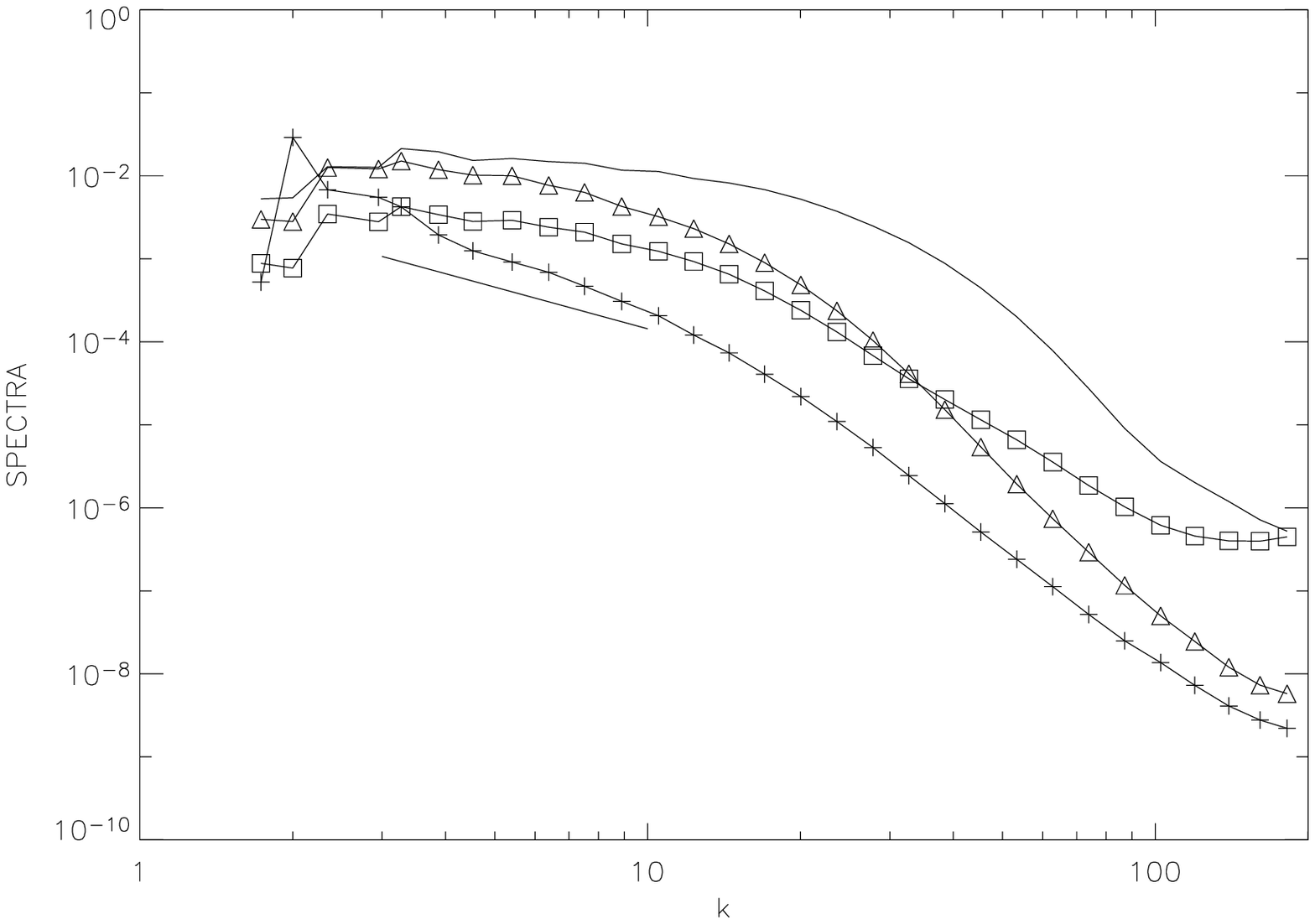}
\hskip 0.08\textwidth 
  \caption{ {\it left panel}: Mach number is 0.35,
 {\it central panel}: Mach number is 7 (figures are from CL03),
{\it right panel}: Mach number is 7, power spectra
 of: {\it solid} -- density, {\it crosses} -- velocity, {\it triangles}
-- logarithm of density, {\it squares} -- density restricted from above
by $2\rho_0$, short streak is a Kolmogorov scaling, $k^{-5/3}$.}
\label{fig_coupling}

\end{figure*}

\section{What is the scaling of fast waves?}

At both low and high $\beta$ fast magnetosonic waves have nearly
isotropic dispersion relation, which makes them similar to simple
acousic waves. In fact, the analytical dependence of the wave speed
on the angle between ${\bf k}$ and ${\bf B}$ is the same for the values
of $\alpha=c_S^2/V_A^2=\beta(\gamma/2)$ equal to $x$ or $1/x$. So
the case of $\alpha=1$ represents most anisotropy possible.
In the so called wave turbulence approach
perturbations of physical quantities are represented as a combination
of weakly interacting waves. (Zakharov, 1965; Zakharov, Sagdeev, 1970;
Kats, Kontorovich, 1973)

In the leading order three wave processes waves obey conservation laws
$\omega=\omega_1+\omega_2$, $\bf{k}={\bf k_1}+{\bf k_2}$. The form of resonant surfaces
that came from solving above equations play crucial role in the nature
of wave interactions. For a nondispersive acousic wave $\omega=ck$ and
the resonance occurs only when all three ${\bf k}$ are collinear.
It could be shown that despite the anisotropy, even for the case of
$\alpha=1$ fast MHD waves can interact only along rays. In this
sense both sound and fast waves are a special case in the theory
of weak or wave turbulence.

The decomposition into interacting waves is conducted by
the change to normal variables $a_i(k)$, $a_i^*(k)$,
which are the classic versions of creation and
annihilation operators, and leaving only quadratic and qubic terms
of the power series of the Hamiltonian with respect to $a$
This method could not be questioned until the
perturbation amplitudes are small, however, usually,
wave turbulence approach takes another step of assuming
the randomness of phases, which allows to write the the kinetic
equation
for the $\langle a_k a^*_k \rangle$, or the occupation numbers
of the waves.
The applicability of this approach is seriosly
questioned in the case of nondispersive waves and collinear
interaction, since waves, travelling in the same direction with
the same speed have an infinite time to interact, hence their
phases should reveal growing deviation from randomness. This
is clearly demonstrated on the analythically solvable one-dimentional
Burgers equation. In this case shocks are formed after finite time,
even with arbitrarily small nonlinearity. Shocks could be described
as a collection of waves with specific phases.

However the validity of the kinetic equation approach has been
advocated recently in 3D in the framework of the generalized kinetic
equation, in which waves are slightly damped (L'vov, et al. 2000).
In the above paper slight defocusing of isotropisation of the distribution
function reported and the damping is estimated from the nonlinear
interaction itself.

The kinetic equation approach is thought to work much better
in the case of dispersive waves, which allow for noncollinear
three wave interactions. For example, if $\omega=k^{1+\epsilon}$, 
$\epsilon>0$, triangle inequality allows for such interactions,
while for $\epsilon<0$ it doesn't. It was shown (L'vov, Falkovich, 1981)
that the above dispersion with $\epsilon>0$ lead to focusing, i.e.
anisotropic part of the distribution increases faster with k
than isotropic one. This effect could be understood from the
simple physical picture of the resonantly interacting three
waves with ${\bf k}$ being the sum of ${\bf k}_1$ and ${\bf k}_2$
and $k<k_1+k_2$. The angle between ${\bf k}_1$ and ${\bf k}_2$
is in most cases larger than the angle between ${\bf k}$ and ${\bf k}_1$
or ${\bf k}_2$.

Even in the case of dispersive waves the validity of kinetic equation
approach has been questioned (see, e.g. Majda et al. 1997) in  1-dimensional
case by numerical simulations. In this case various types of structures
that violate the randomness of phase has been proposed (Zakharov, et al, 2004)


\def\eps{\epsilon}
\def\dd{\partial}

Self-similar solutions for several types of waves were constructed
in the kinetic equation approach using Zakharov transformations
in ${\bf k}$ space in the collisional integral. (Zakharov, 1965;
Katz, Kontorovich, 1973). For the particular case of the acoustic wave
turbulent 1D energy density was predicted to scale like $k^{-3/2}$.

\section{What is the scaling of density?}

Density at low Mach numbers follow the GS95 scaling when the
driving is incompressible (CL03). However, CL03 showed that
this scaling substantially changes for high Mach numbers.
Fig.~4 shows that at high Mach numbers
density fluctuations get isotropic. Moreover, our present
studies confirm
the CL03 finding that the spectrum of density gets substantially
 {\it flatter} than the GS95 one (see also Cho \& Lazarian 2004).
 Note, that a model of random shocks would produce
a spectrum {\it steeper} than the GS95 one.

The study of the density perturbation scaling properties
and statistics regarless of the perturbation origin
(slow or fast modes) in numerical simulations or observations
could be a model-independent test for applicability of the
incompressible MHD approach in small scales. Indeed,
incompressible MHD requires divergent-free motions, and
set the density constant everywhere. Flatter density spectrum
has been reported in both low-beta MHD and purely
hydrodynamic high Mach simulations.

In the highly compressible case we expect significant
non-linear perturbations of density, but what the nature
of these perturbations? In a compressible isothermal
hydrodynamics equations are invariant under transformation
$\rho \rightarrow C\rho$, it is, then, natural to consider
$\log\rho$ so that equations have additive symmetry.
If we expect the log-density buildup to be a random process,
then a PDF of log-density will be gaussian. It was confirmed
in high-Mach hydrodynamic 1D simulations. For the arbitrary
polytropic equation of state $\gamma\not=1$ the PDF developed
power-law tails on both sides. (Passot, Vazquez-Semadeni, 1998)

Our figure 4 confirmes that the flattening of the density spectra 
3D simulations is
due to very high density spikes, normally present in both
MHD and purely hydrodynamic data. Indeed, both the
spectra of log-density and the spectra of density, constrained
from above by several $\rho_0$ do not have flattening, but
bear good resemblance to the velocity spectra, and have
scaling close to Kolmogorov. A simple removal of high density
peaks also results in the Kolmogorov-type spectrum, which
confirms the origin of peaks due to high density clupms
produced by large scale driving. 

It should be noted that this result holds true even in the
case of relatively highly magnetized fluid, with $M_A$ of
0.5. It is natural to suggest then that a random buildup
of log-density, which in magnetized case will be governed
by slow mode and be essentially 1-dimensional, works as well
in this case, even though MHD equations no longer
posess the symmetry of $\log\rho \rightarrow \log\rho+C$.

The result that a highly compressible turbulence leaves
the density field perturbed by 2-3 orders of magnitude,
and even higher in small areas might undermine certain
models of incompressible turbulent transport, that require
weak interaction of waves making transport non-local.

\section{Is MHD turbulence and Hydro turbulence similar?}

Not only Kolmogorov scaling indicate the similarity 
between the hydro and MHD turbulence. Studies of higher
order correlations when the motions perpendicular to 
the {\it local} magnetic field are considered show high
degree of similarity between the magnetized and non-magnetized
turbulence (CLV02b). This is suggestive that motions perpendicular
to local magnetic field are essentially hydrodynamic. For some
problems, e.g. turbulent heat transport (see Cho et al 2003),
this entails similar results with and without ${\bf B}$.

Nevertheless, the most important difference between hydro and
magnetic turbulence is the {\it scale-dependent}
anisotropy of the latter. This peculiar type of anisotropy
entails dramatic consequencies for the transport of cosmic rays
or acceleration of charged dust (see Yan \& Lazarian 2004 and
references therein).

\section{Summary}
~~~~1. 
MHD turbulence is not a mess. The turbulent cascade consists of
Alfven, slow and fast modes cascades.
Fast modes follow accoustic turbulence cascade, while Alfven and
slow modes are similar to their counterparts in incompressible
MHD.

2. Fast decay of MHD turbulence is not due to strong coupling of
compressible and incompressible motions. The transfer of
energy from Alfven to compressible modes is small. The Alfven mode
develops on its own and decays fast.

3. Density fluctuations follow the scaling of Alfvenic part of the
cascade only at small Mach numbers. At large Mach numbers the log-density
shows Kolmogorov spectrum.

{\bf Acknowledgments}{\it 
We are grateful to Jungyeon Cho for data and nice discussion.
We acknowledge NSF grant AST 0307869 and the NSF Center
for Magnetic Self-Organization in the Laboratory and Astrophysical Plasmas}.

%

\end{document}